\pdfoutput=1

\documentclass[11pt]{article}
\usepackage{moriond,epsfig}
\usepackage[latin1]{inputenc}
\usepackage[OT1]{fontenc}

\def\Journal#1#2#3#4{{#1} {\bf #2}, #3 (#4)}

\def\NIMA{{\em Nucl. Instrum. Methods} A}

\def\PRD{{\em Phys. Rev.} D}

\def\be{\begin{equation}}
\def\ee{\end{equation}}
\def\bea{\begin{eqnarray}}
\def\eea{\end{eqnarray}}

\begin{document}
\vspace*{4cm}
\title{QUIET - MEASURING THE CMB POLARIZATION WITH COHERENT DETECTOR ARRAYS}

\author{ D. SAMTLEBEN \footnote{on behalf of the QUIET collaboration {\it quiet.uchicago.edu}}}

\address{Max-Planck-Institut f\"ur Radioastronomie, Auf dem H\"ugel 69,\\
53121 Bonn, Germany}

\maketitle \abstracts{The footprint of inflation in the polarization
pattern of the CMB is expected at more than an order of magnitude
below the limits of current polarization measurements. Large receiver
arrays are mandatory for achieving the required sensitivity. We
describe here the approach of the Q/U Imaging ExperimenT (QUIET)
which will use coherent detector arrays. Pseudo-correlation
receivers have been produced at 40 (90) GHz in small massproducable
chip packages. Deployment of the first arrays with 19 (91) receivers
is taking place in 2008 in the Atacama Desert in Chile and an expansion
to $\sim$1000 receivers is foreseen for the future. The two
frequencies and the selection of observing regions with minimal
foreground and good overlap with other upcoming experiments will enable
the detection of tensor to scalar ratios $r\sim10^{-2}$.}
\section{Introduction}
Measurements of the Cosmic Microwave Background Radiation helped
significantly in establishing the current cosmological model
\cite{wmap}. The CMB temperature anisotropies have already been
measured to a high precision, while only in recent years experiments
achieved sensitivities to detect and measure the much smaller
polarization anisotropies. The polarization pattern is divided into
the E- and B-modes where the former derive from the primordial density
fluctuations and the latter from primoridal gravity waves and from the
lensing of the E-modes by matter in the line-of-sight. The yet
undetected B-mode signal from primordial gravity waves can offer
unique insights into the inflationary period \cite{infl}. Its size is
model-dependent, parametrized by the tensor-to-scalar ratio $r=T/S$,
but it is expected to be at least one order of magnitude smaller than
the E-modes so that a significant increase in sensitivity is required
of future experiments. Since current receivers are nearing fundamental
limits this leap in sensitivity is only feasible with large receiver
arrays.
\section{Status of Measurements}
\begin{figure}
\vspace*{-3cm}
\hspace*{-1cm}\includegraphics[angle=90,width=14cm]{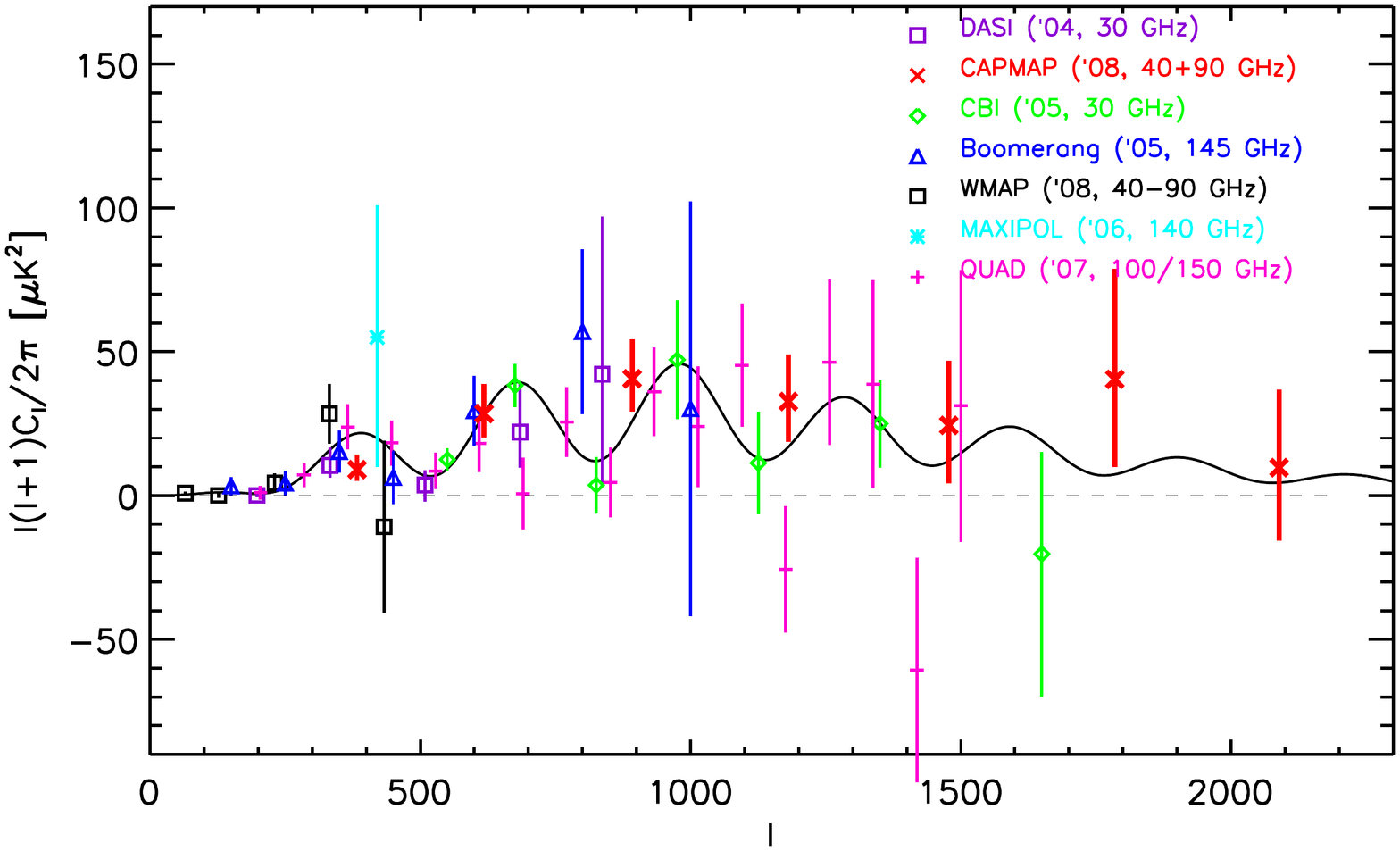}
\caption{Current measurements of the E-mode power spectrum on top of the concordance model. \label{fig:status}}
\end{figure}

Figure \ref{fig:status} shows the status of the E-mode masurements
from which it becomes clear that much higher sensitivity is required
for meaningful constraints of the cosmological model. The first CMB
polarization measurements were all performed by coherent detectors
(DASI, CBI, CAPMAP, WMAP) while by now also bolometric experiments
have proven their polarization capabilities for the CMB measurements
(Boomerang, QUAD) with significant detections. Since the bolometers
offer high sensitivity together with the option of an easy scaling to
large arrays they are considered the main path for future CMB
polarization experiments. However, both coherent and bolometric
systems are crucial in approaching the challenge of measuring the tiny
B-mode signal, so that the results can independently be verified by
methods with different systematics.  The most recent CMB polarization
results from ground-based coherent detectors came from the CAPMAP
experiment \cite{capmap}. 12 W-band (90 GHz) and 4 Q-band (40 GHz)
receivers were used on the Crawford Hill 7~m antenna in New Jersey.
Data from $\sim$1660 hours of observing a 7.3 (9.2) square degree
region around the NCP in W(Q)-band were processed in two independent
pipelines. The final power spectra were evaluated from a selection of
$\sim$950 hours with consistent results from both pipelines. A series
of null tests ensured the absence of contamination from ground-signals
or foregrounds. The measurement is competitive with the other
ground-based efforts and extends the measurements to higher
multipoles. The experiments to date have used at most a few 10s of
receivers and therefore they were only able to provide upper limits
for B-modes.
\section{The Q/U Imaging ExperimenT (QUIET)}
The developments for the low frequency receivers of the Planck space
mission at the Jet Propulsion Laboratory (JPL) led to the integration
of a pseudo-correlation receiver in a compact chip package, allowing
for a mass production of coherent receiver arrays at low cost
\cite{quiet}. The QUIET collaboration formed in 2003 around the groups
involved in the CMB polarization experiments CAPMAP and CBI to take
advantage of these receivers for measurements of the CMB
polarization. By now it consists of experimental groups from 12
institutes (Max-Planck-Institut f\"ur Radioastronomie Bonn, Caltech,
Columbia University, JPL, Kavli Institute for Cosmological Physics at
the University of Chicago, Kavli Institute for Particle Astrophysics
and Cosmology at the Stanford University, KEK, University of
Manchester, University of Miami, University of Oslo, University of
Oxford, Princeton University). Two prototype detector arrays are being
built, one with 91 elements at 90~GHz and one with 19 elements at
40~GHz.

\begin{figure}[t]
\vspace*{-1cm}
\includegraphics[width=5cm]{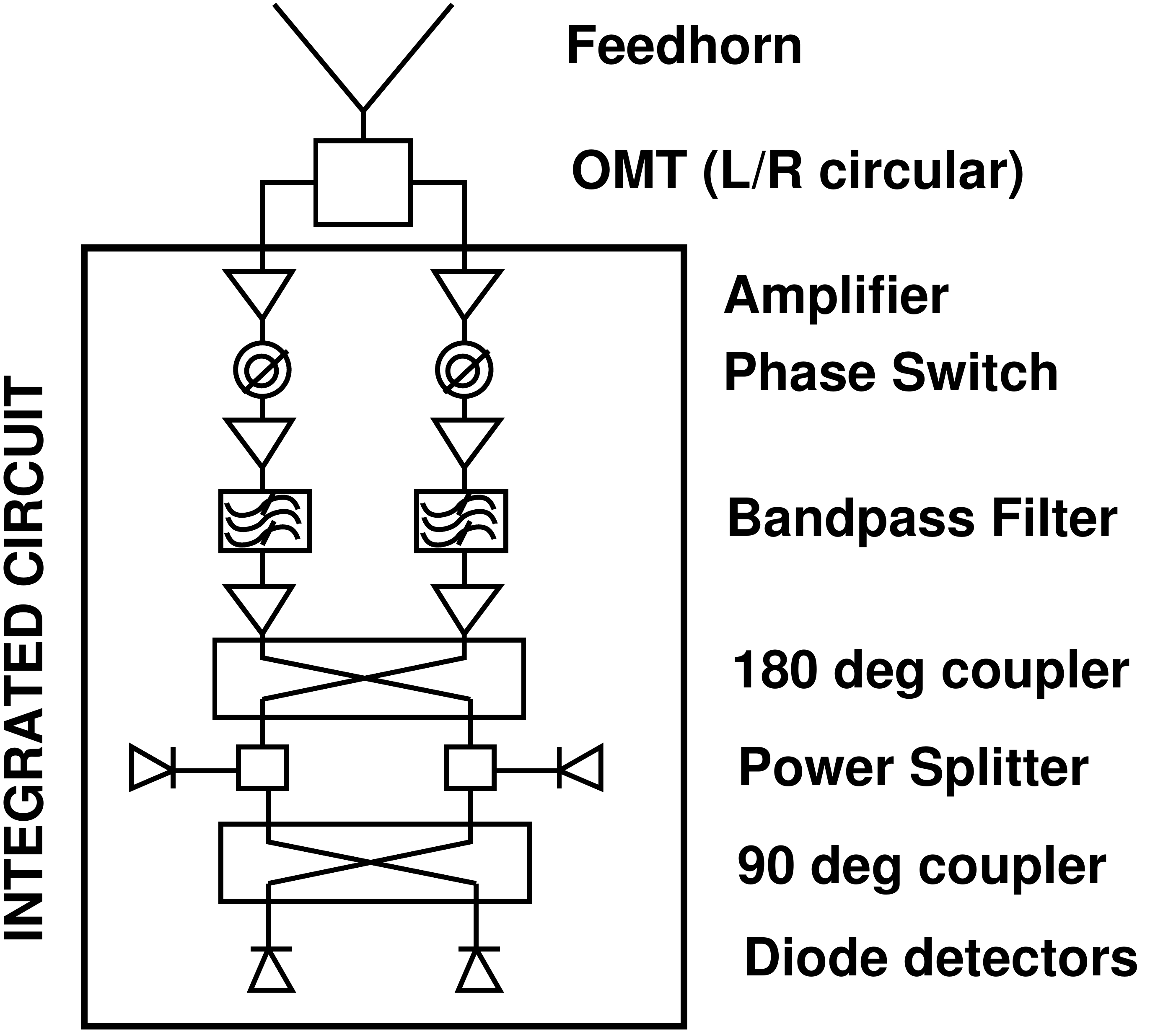}
\hspace*{2cm}\includegraphics[width=6cm]{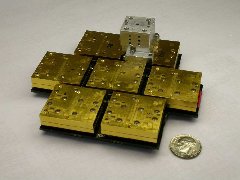}
\caption{Left: Module sketch. Right: Subarray of 7 W-band modules together
with one OMT.
\label{fig:module}}
\end{figure}

The scheme of the receiver together with a photo of the W-band
(90~GHz) prototypes can be seen in figure \ref{fig:module}. The
incoming radiation is coupled via a feedhorn to an Orthomode
Transducer (OMT) which then feeds the chip receiver with the left and
right circularly polarized components. The feedhorns were produced as
platelet arrays, using about 100 plates with different hole patterns which
when bonded together form a corrugated horn array. The high bandwidth
(20$\%$) OMTs were produced as septum-OMTs in split-block technique.

Within the receiver chip package the radiation is amplified by Indium
Phosphide (InP) High Electron Mobility Transistors (HEMTs) implemented
as Monolithic Microwave Integrated Circuits (MMICs). In one of the
legs the signal is phase switched at 4~kHz which is well above the 1/f
knee of the amplifiers of $\sim 1~$kHz. After passing through a
bandpass filter and another amplifier stage the signals from the two
legs are combined in a 180 degree hybrid and coupled to diodes. The
demodulated signal on those diodes corresponds to the Stokes parameter
Q. A power splitter couples a part of the radiation via a 90 degree
combiner to two additional diodes, where the demodulated signal
corresponds to the Stokes parameter U. Measuring Q and U simultaneously, which allows valuable 
systematic checks, is unique to coherent technology.

The receivers together with the OMTs and horns are mounted in a
cryostat which is cooled by a closed-cycle helium fridge to 20~K to
ensure low noise from the amplifiers.

The signal from the diodes is led out of the cryostat where it is
digitized on a custom electronics board by 800~kHz ADCs and
demodulated using a Field Programmable Gate Array (FPGA) \cite{daq}. The 
digital demodulation allows a flexible processing of the data stream, the
monitoring of high-frequency noise and the creation of quadrature
demodulation data sets for a check of systematics.

The averaged signal on the diodes gives a measure for the total power
but is compromised by the 1/f noise of the amplifiers since no phase
switching is applied. In order to retrieve clean total power
information 10~$\%$ of the receivers were modified for a differential
total power measurement: The orthogonal polarizations from the 
signal of neighboured feeds are coupled via a Magic Tee to the chip 
packages so that the demodulated output on the Q diodes provides a 
measurement of the temperature difference measured between the two feeds. 
This addition is crucial for CMB TT and TE measurements and for the 
identification and removal of unpolarized foregrounds.

The cryostat together with the primary and secondary mirror of a new
1.4~m telescope is surrounded by a box covered with eccosorb to form
an absorbing ground screen. The structure will be mounted on the CBI
platform in the Atacama Desert in Chile (5080m) where the high
altitude and the dry atmosphere provide excellent observing conditions
with little atmospheric contamination.

The 40~GHz array is undergoing final tests and is scheduled to be
shipped to Chile in June 2008. The 90~GHz array is still in
production and planned to be finalized by the end of 2008. The work on
the possible extension of the arrays to up to $\sim$1000 receivers is
in progress.
\section{Science reach}
QUIET will focus its sensitivity on four selected patches with minimal
foreground contamination, each 400 square degrees. The exact
observing regions will be coordinated with other experiments with
different frequencies (Polarbear, Clover, Ebex) in order to allow
optimal foreground characterization and removal.

\begin{figure}
\vspace*{-11cm}
\hspace*{-2cm}\includegraphics[width=12.8cm]{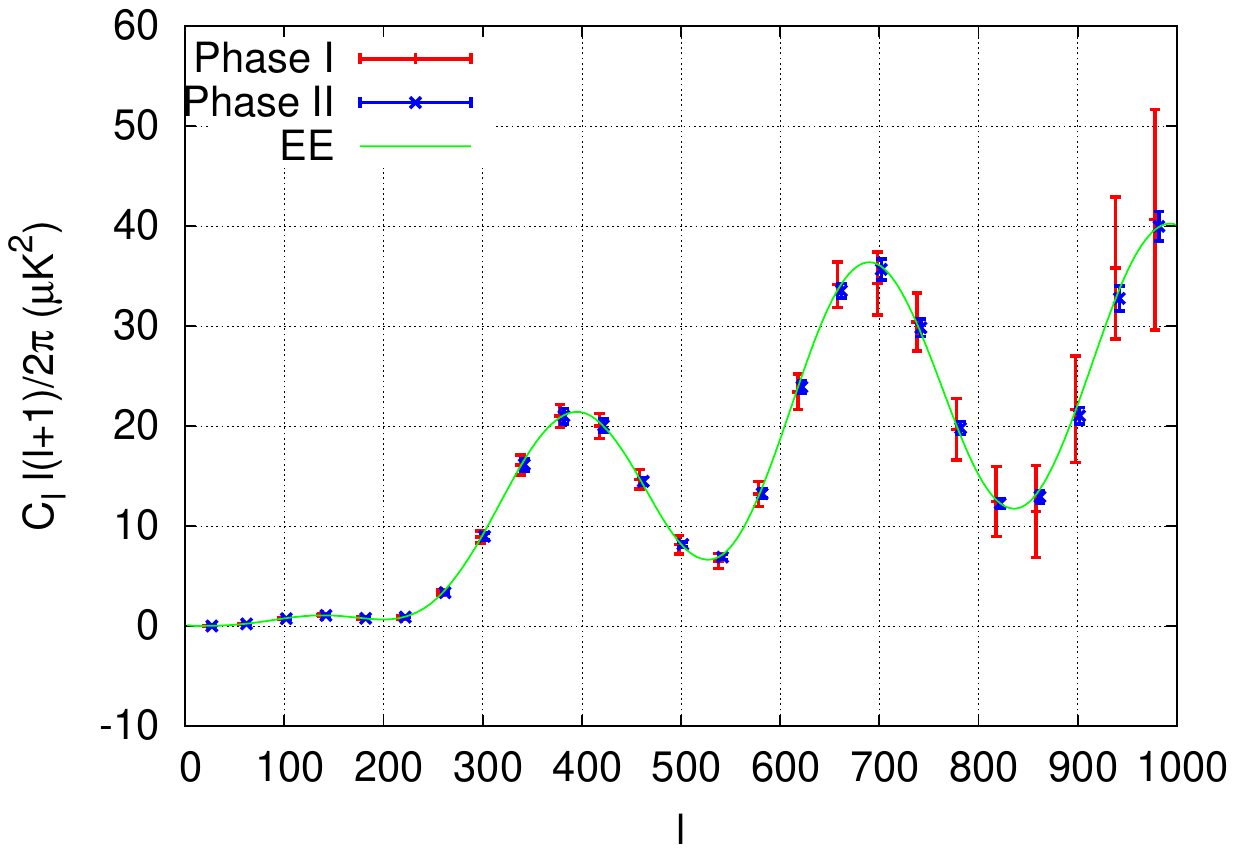}
\hspace*{-5cm}
\includegraphics[width=12.8cm]{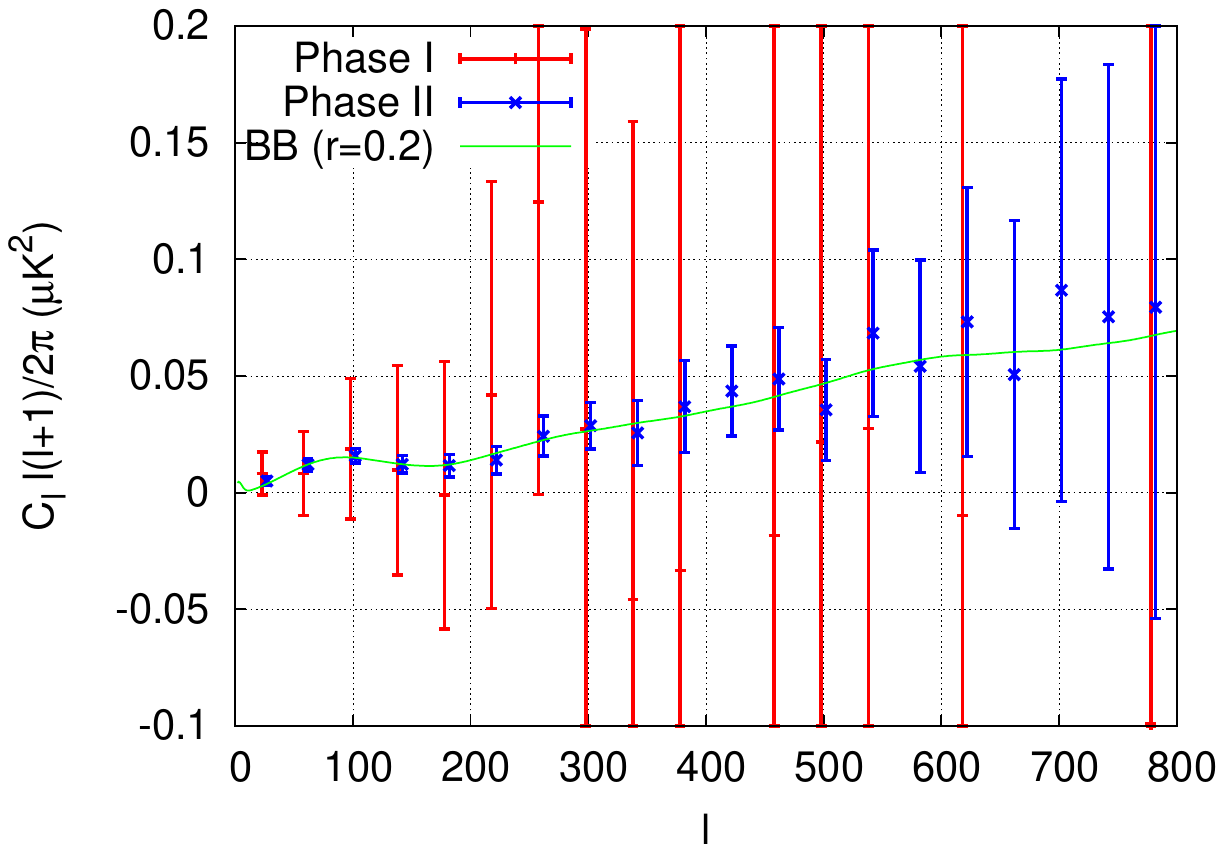}
\caption{Left (Right): Expected results for the E-mode (B-mode) power spectrum.
\label{fig:powspec}}
\end{figure}

A first analysis pipeline using time-stream filtering for 1/f removal,
optimal map-making and the Pseudo-$C_l$ method \cite{pseudo} for the
power spectrum estimation is being developed and already in use for
the investigation of several systematics. The projected measurement
errors for the E- and B-mode power spectra from this pipeline are
shown in figure \ref{fig:powspec}. The error bars for phase I (II) are
derived using the expected sensitivity of a 50 (500) element
W-band array observing for 10 (20) months at 50$\%$ efficiency. In
order to account for the sensitivity loss from foreground removal the
Q-band sensitivity has not been used for these estimates.

In phase I the first peaks of the E-mode spectrum will be measured to
unprecedented precision. Expanding the arrays will give access to the
B-modes from gravitational waves. QUIET will be able to give a
comparable precision on cosmological parameters other than $r$ as
expected from the space mission Planck but provide deeper maps of
small regions of the sky.  While Planck will be able to measure a
tensor-to-scalar ratio of $r=0.05$ using the reionization signal at
very low $l$ QUIET will target the maximum of the B-mode signal at
$l=100$ and be able to reach $r \sim 10^{-2}$.
\section{Conclusion}
The first coherent detector array at 40~GHz will start data taking
in mid 2008. With a 90 GHz array following in the winter the 
sensitivity will allow measurements of the E-mode spectrum at an 
unprecedented precision. Expansions of the prototype arrays to up to 1000 
detectors are foreseen and will provide the sensitivity to access 
the B-modes from primordial gravity waves down to levels of $r \sim 10^{-2}$.
\section*{References}

\end{document}